\documentclass[conference,a4paper]{IEEEtran}
\IEEEoverridecommandlockouts
\usepackage{cite}
\usepackage{amsmath,amssymb,amsfonts}
\usepackage{algorithmic}
\usepackage{graphicx}
\usepackage{textcomp}
\usepackage{xcolor}
\def\BibTeX{{\rm B\kern-.05em{\sc i\kern-.025em b}\kern-.08em
    T\kern-.1667em\lower.7ex\hbox{E}\kern-.125emX}}

\usepackage{subfigure}
\usepackage{fancyhdr}
\usepackage{enumitem}
\usepackage{algorithm}
\usepackage{algorithmic}

\begin{document}

\title{Universal Non-Intrusive Load Monitoring (UNILM) Using Filter Pipelines, Probabilistic Knapsack, and Labelled Partition Maps
\thanks{Research supported by NSERC Discovery Grant RGPIN-2018-06192.}
}


 \author{\IEEEauthorblockN{Alejandro Rodriguez-Silva}
 \IEEEauthorblockA{\textit{School of Engineering Science} \\
 \textit{Simon Fraser University}\\
 Burnaby, Canada \\
 Email: ara61@sfu.ca}
 \and
 \IEEEauthorblockN{Stephen Makonin}
 \IEEEauthorblockA{\textit{School of Engineering Science} \\
 \textit{Simon Fraser University}\\
 Burnaby, Canada \\
 Email: smakonin@sfu.ca}
 }

\maketitle

 \thispagestyle{empty}
 \renewcommand{\headrulewidth}{0.0pt}
 \thispagestyle{fancy}
 \lhead{}
 \chead{This paper is currently under review for an IEEE conference.}
 \rhead{}
 \lfoot{}
 \cfoot{Copyright \copyright{ } 2019 the authors.}
 \rfoot{}

\begin{abstract}
Being able to track appliances energy usage without the need of sensors can help occupants reduce their energy consumption to help save the environment all while saving money. Non-intrusive load monitoring (NILM) tries to do just that. One of the hardest problems NILM faces is the ability to run unsupervised -- discovering appliances without prior knowledge -- and to run independent of the differences in appliance mixes and operational characteristics found in various countries and regions. We propose a solution that can do this with the use of an advanced filter pipeline to preprocess the data, a Gaussian appliance model with a probabilistic knapsack algorithm to disaggregate the aggregate smart meter signal, and partition maps to label which appliances were found and how much energy they use no matter the country/region. Experimental results show that relatively complex appliance signals can be tracked accounting for 93.7\% of the total aggregate energy consumed.
\end{abstract}


\section{Introduction}

Disaggregation is a difficult, ill-posed problem that uses statistical models and algorithms to determine the unknown components that were used to sum the known aggregate value. 

\begin{figure*}[th]
  \centering
  \includegraphics[width=\textwidth]{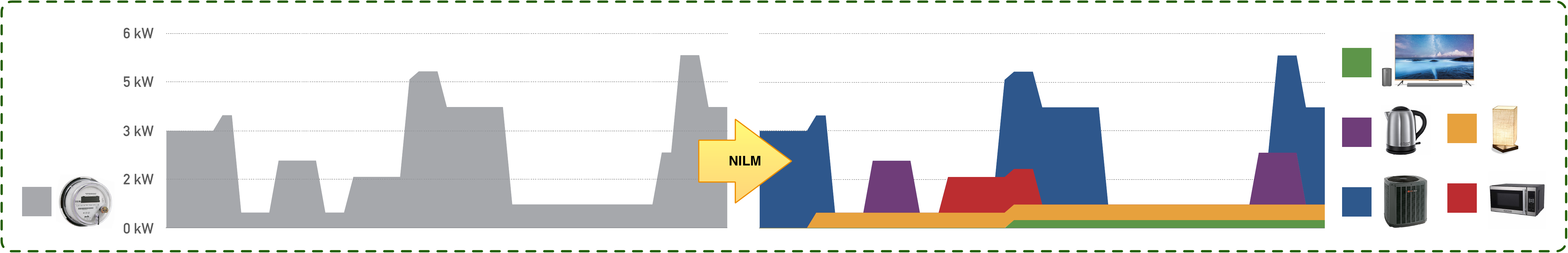}
  \caption{Sensing appliance power usage and consumption without sensors.}
  \label{fig:example}
\end{figure*}

Disaggregating power/energy data is known as \textit{non-intrusive load monitoring} (NILM) --- using a building's smart meter to track or sense appliance usage (see Figure~\ref{fig:example})~\cite{Hart_1992}. 
NILM uses computation, without the intrusion or cost of sensors, to model and understand the power consumption of loads/appliances in houses or buildings with the goal of reducing energy consumption.
Appliance-level data inferred by NILM can be used by occupants to help make informed choices (that fit their lifestyle) as to how they want to conserve. 
Ultimately this is a win for their pocketbook and the environment.

\subsection{Mitigation by Conservation}

In 2012, Natural Resources Canada (NRCan) reported that Canadian households accounted for 14.6\% of total energy-related greenhouse gas emissions. 
Despite various government incentives and more energy-efficient appliances on the market, Canadian households have increased their energy consumption by 4.5\% in 2013 (97.5 GJ/household) from 2011 (93.3 GJ/household). 

A recent study \cite{Chakravarty_2013} showed that 80\% of participants want to have access to disaggregation data (i.e. knowing how their appliances consume energy) and believed that everyone should have access to this information. 
This study also shows when load disaggregation information is made available to occupants, those occupants can reduce their energy consumption by an average of 14\% by changing appliance use habits. 
In fact, more research is showing that in order for occupants to reduce (on average) their energy consumption by more than 9\%, real-time appliance-specific consumption information is needed \cite{Ehrhardt_2010,Armel_2013}. 

If we look at the COP21 Paris Climate Agreement, Canada\rq{}s commitment is to cut carbon emissions by 30\% below 2005 levels by 2030 which is similar to that of the USA. 
If we solve the difficult problem of NILM in five years, we will have the means to reduce the amount of carbon emissions possible for homes and buildings with ample time to deploy it. 

As an example, we will use numbers from the USA Environmental Protection Agency (EPA) as they are readily available and contain data as recent as 2015.
In 2015, EPA reported a 12\% reduction in carbon emissions, to date. 
This leaves 16\% still needing to be reduced (or 960Mt of carbon).
The EPA reports households (including commercial) emit 847Mt of carbon/year, which represents 12\% of the total annual emissions. 
This means that from this economic sector a further 115Mt of carbon would need to be reduced to meet USA targets.
We know that households can reduce their energy consumption by an average of 14\% with real-time appliance-specific feedback.
This 14\% reduction is equivalent to 118.6Mt of carbon/year which is beyond the 115Mt reduction needed.
This means the reduction goal set for the household and commercial economic sector has been met.

\subsection{Cost Benefit Case Study}

\begin{figure*}[th]
  \centering
  \includegraphics[width=\textwidth]{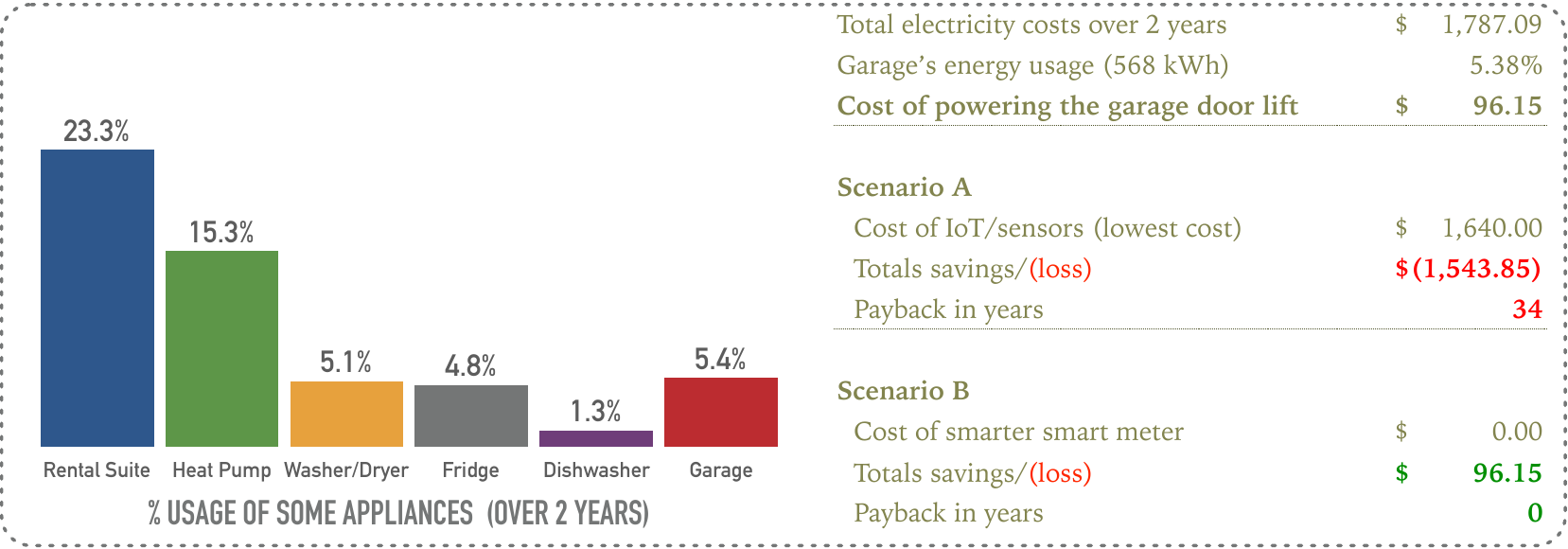}
  \caption{Results (left) of a two-year case study using AMPds~\cite{ampds2} where a 1980's wireless garage door lift was found to have consumed a large amount of energy. The garage was only used for storage -- the lift did not even need to be powered. Analysis (right) shows that using sensors (Scenario A) would cost far more than the amount of energy savings realized \textit{vs.} using NILM (Scenario B) where savings are realized immediately. Costs are based on \$0.08/kWh.}
  \label{fig:case}
\end{figure*}

If we can provide appliance-specific consumption information using NILM, then we can make use of the digital ``smart'' meters installed on almost every household in Canada. 
Figure~\ref{fig:case} shows a two-year study that compares the cost of equipping a house with sensors \textit{vs.} a NILM system.

To realize a small saving in energy conservation would require decades of savings (in this case 3.5) to payback to cost of purchase and installation of IoT sensors.
Note, this cost is conservative as it only takes into account just the cost of 24-meter sensors and data logger --- not the cost of having a professional electrician install the sensors and logger, nor does it take in account the energy costs of run the sensors and data logger continuously.

Many households would find the cost of these sensor systems unaffordable resulting in an adoption barrier.
With a NILM system, energy savings can be realized immediately without an investment in sensors.

\subsection{Research Contributions}

As a result of this work we have made the following contortions and provided a state-of-the-art NILM algorithm.
\begin{enumerate}
    \item We show why NILM to a useful problem to solve to reduce energy demand and reach sustainability goals.
    \item We demonstrate how a filter pipeline can create a clean, sharp signal.
    \item We have designed a NILM algorithm that does not require prior training (nor uses general appliance models).
    \item We show how using a labelled partition map can allow disaggregation independent of country/region with very different appliance mixes and operational characteristics.
\end{enumerate}

\section{Background}

The first peer-reviewed NILM publication was in 1992~\cite{Hart_1992} and with the recent release of publicly available datasets~\cite{kolter2011redd,ampds2,refit,makonin2018rae} there is a resurgence in interest in solving NILM. 

It is clear that at best NILM is a semi-supervised problem as labelled data is needed to properly identify what appliances have been disaggregated. Setting aside the labelling problem, the disaggregation part of NILM can be potentially solved via an unsupervised learning algorithm. Thus, Unsupervised NILM refers to this part and sets aside the labelling problem as something to be solved separately, perhaps through some additional algorithm or occupancy feedback system.

Supervised learning solutions are proven to be accurate~\cite{7317784}, but brittle, making them undeployable in a real world situation. Although, deep learning methods~\cite{8683552,8682860,8682543} may show promise, if enough data is available. As a result, all serious algorithmic-based NILM research is now focused on an unsupervised NILM solution.

In terms of using advanced filters for preprocessing data there has been some work done. However, only one advanced filter is ever used~\cite{8340730,zhao2018improving}.

\subsection{General Model Tuning}

Some NILM algorithms~\cite{PARSON20141,murray2019trans} require priors to build a general model then tune it to a specific house. Some success with houses in the same country/region. However, this has not been proven to be successful inter-country/region; e.i., having a model of a dishwasher in the UK will not work for disaggregating dishwasher in USA. This method is often referred to as transfer learning~\cite{murray2019trans}.

\subsection{Online Learning}

The goal of online learning is to discover and create appliance models without the use of priors. This is an ideal solution as it requires no training and can adapt to different appliance mixes and different operational characteristics that are country/region dependent. There has not been a concerted effort to solve NILM in this fashion as it would be considerably harder to do when compared to a transfer learning solution. Our proposed UNILM method attempts to solve NILM using this approach by combining the probability of Gaussian distributions and the optimization of the Knapsack Problem. 
Using Knapsack for NILM has been looked at before, albeit as part of a larger genetic algorithm solution~\cite{egarter2013evolving}.

\section{Methodology}

Our proposed solution is considered \textit{universal}. Universal NILM (UNILM) is both unsupervised and transferable. Unsupervised as it relies on no prior trained models as it learns online. Transferable as it can learn appliances in different houses without having prior information about each house.

\subsection{Prepossessing via a Filter Pipeline}

\begin{figure*}[th]
  \centering
  \includegraphics[width=\textwidth]{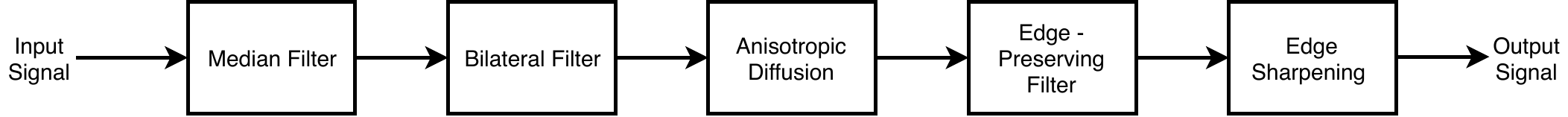}
  \caption{The series of filters used in our proposed filter pipeline. A later version could have filters run in parallel as in typical filter banks.}
  \label{fig:pipeline}
\end{figure*}

\begin{figure*}
\vspace{-0.2in}
    \centering       
        \subfigure{
           \includegraphics[width=0.48\textwidth]{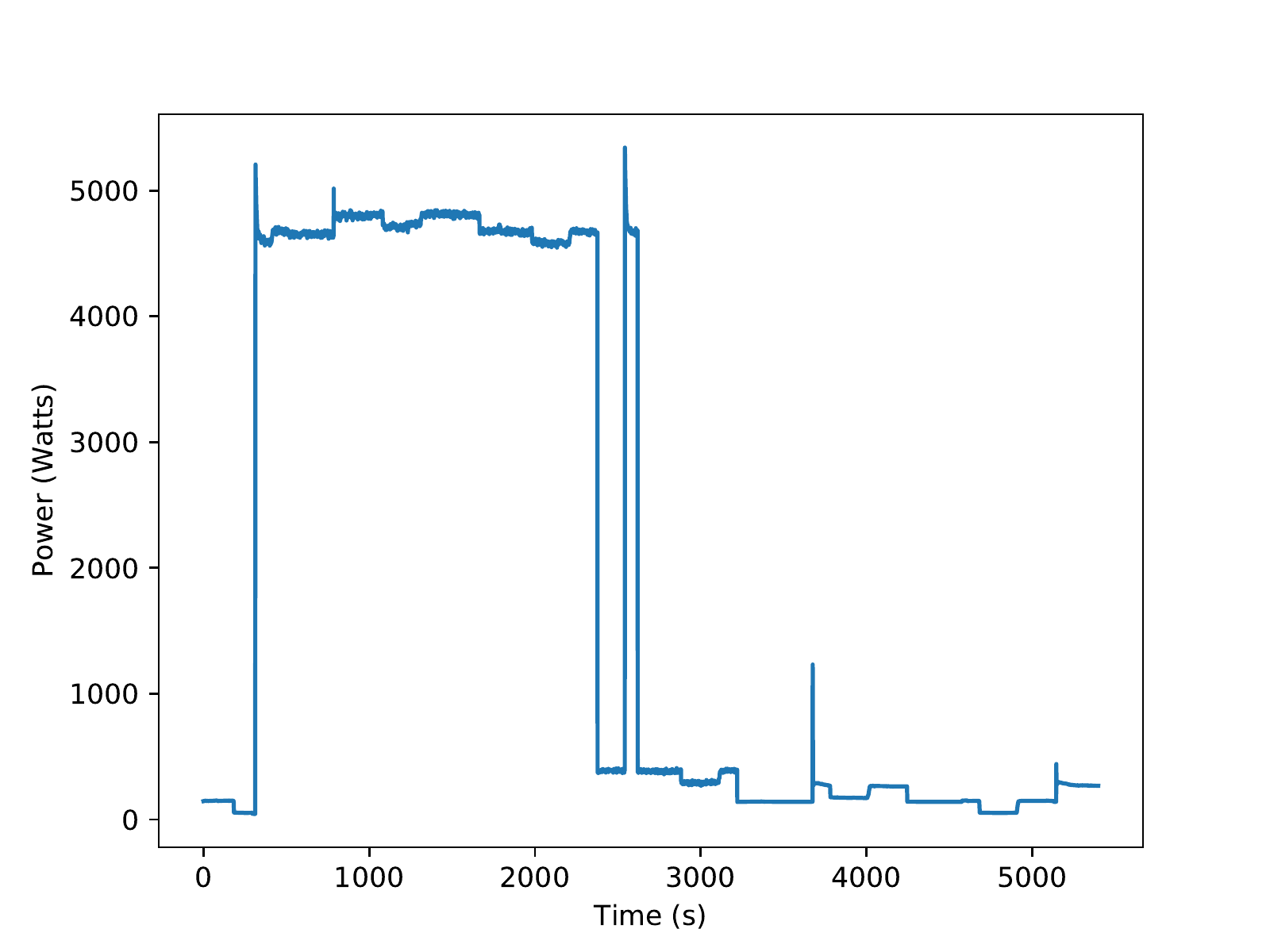}
           \label{fig:unfiltered}
        }
        \subfigure{
            \includegraphics[width=0.48\textwidth]{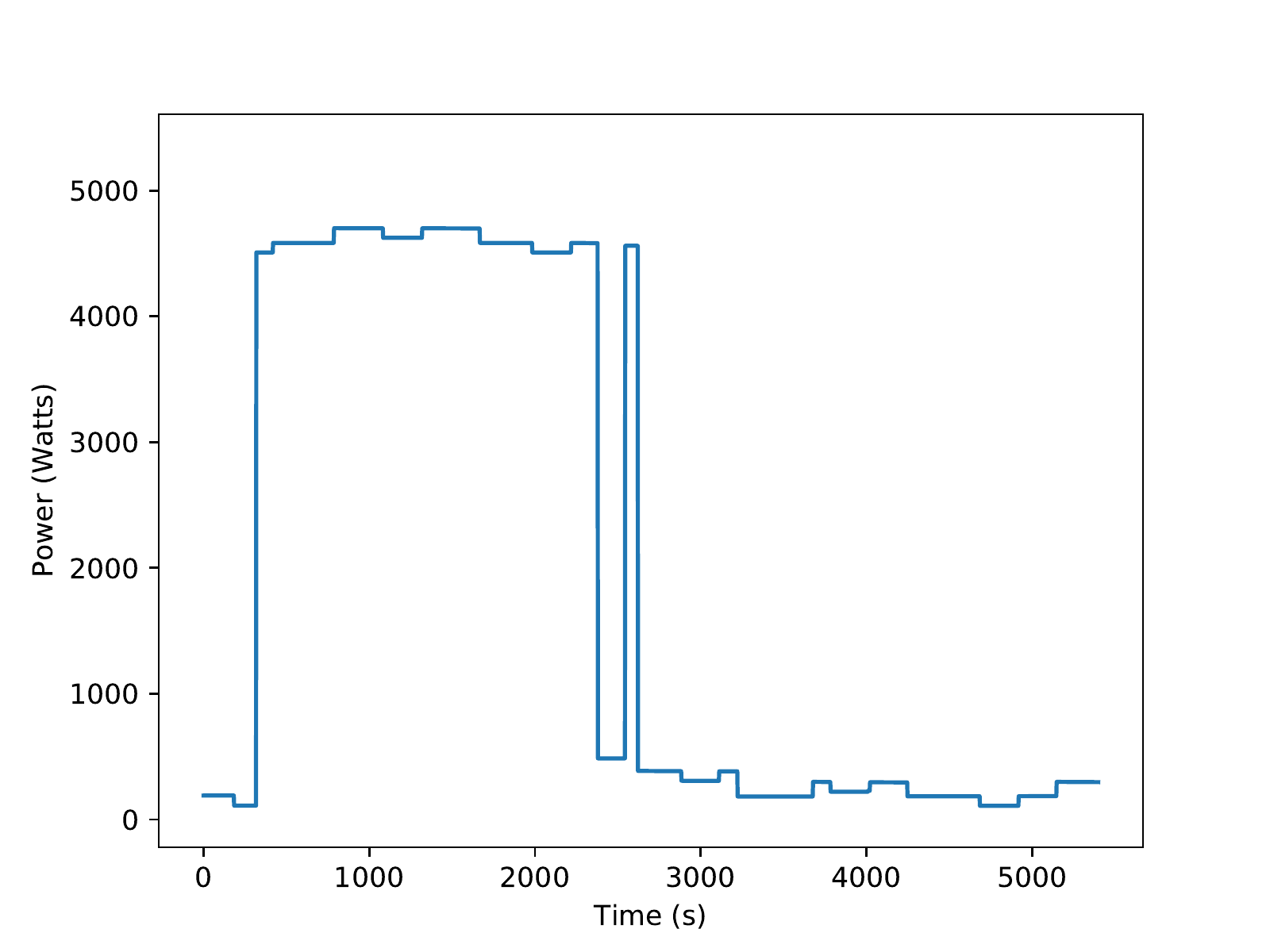}
            \label{fig:filtered}
        }
    \caption{Comparing the raw 1Hz aggregate signal (left) to output of our filter pipeline (right). As expected, clean transients and flat steady-states.}
    \label{fig:filter}
\end{figure*}

We propose using a filter pipeline (see Figure~\ref{fig:pipeline}) for prepossessing the signal as each filter cannot accomplish cleaning a signal with sharp edges and flat steady-states on its own. We use a standard median filter built into Matlab or Python/Numpy (or see~\cite{weiss2012leveraging}) for this first step to remove large spikes (e.g., from fridge compressor start-up). This is followed by sending the signal to a bilateral filter~\cite{tomasi1998bilateral}, then an anisotropic diffusion filter~\cite{perona1990scale}. Next, we use a 2D image processing edge preserving filter that was adapted to process 1D signals as demonstrated in~\cite{gastal2011domain}. In the final step we edge sharpen similar to~\cite{zhao2018improving}. Figure~\ref{fig:filter} shows the results of filtering a 1Hz aggregate power signal.

\begin{figure*}[th]
  \centering
  \includegraphics[width=\textwidth]{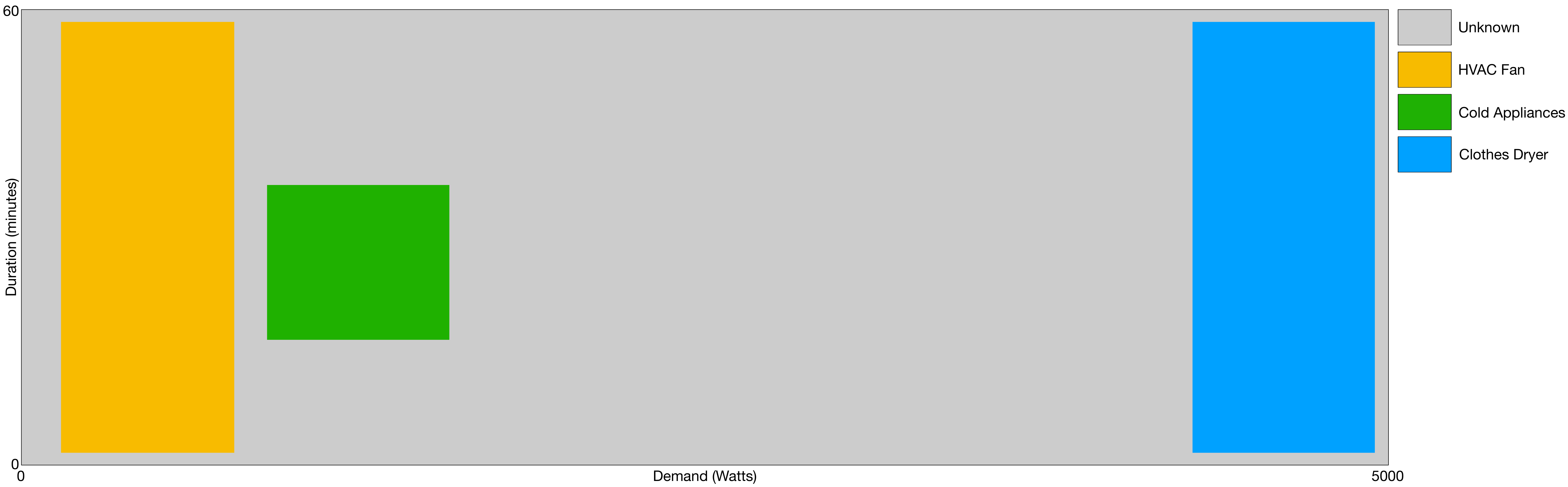}
  \caption{An example of a coloured partition map to auto-label North American appliances. Each country/region would have there own specific partition map based on readily available appliance statistics for government organizations and manufacturers. Using $\mathbf{P}[d, p, r]$, and example where the duration is 30min, the power demand is 4.5kW, and the region is North America (NA); $\mathbf{P}[30, 4500, \text{`NA'}]$  would return the colour ``blue'', which is associated with the appliance label ``Clothes Dryer''.}
  \label{fig:map}
\end{figure*}

We define a filtered signal as: 
\begin{equation}
    y_t = \text{f}(z_t) , 
\end{equation}

\noindent
where $z_t$ is the raw signal at time $t$, $y_t$ is the resulting filtered signal after calling the filter pipeline function. Events are determined by $\Delta y$ which is the derivative of $y_t - y_{t-1}$, where an OFF event is $\Delta y < s$ and an ON event is $\Delta y > s$, else no event has occurred (i.e., seady-state).  Variable $s$ is a minimum threshold step to determine if an event has occurred. For our proposed UNILM algorithm we set to $s = \pm 60$W.

\subsection{Appliance Database \& Probabilistic Knapsack}

We have an appliance database $\mathbf{A} = \{\alpha_1, \alpha_2, ..., \alpha_M \}$ of $M$ appliances found. At time $t_0$ $\mathbf{A} = \varnothing$ (i.e., is empty). Over time as appliances are found, $\mathbf{A}$ will grow in length; therefore, as each new appliance is discovered we have $M = M + 1$.

Each appliance $\alpha_i$ we define a model:

\begin{equation}
    \alpha_i = \{\mathcal{N}_{P0}, \mathcal{N}_{P1}, \mathcal{N}_{D0}, \mathcal{N}_{D1}, \mathbf{y} \}, 
\end{equation}

\noindent
where $\mathcal{N}_{P1}$ is a Gaussian PDF of the ON power demand in Watts, $\mathcal{N}_{D0}$ is a Gaussian PDF of the OFF duration in minutes, $\mathcal{N}_{D1}$ is a Gaussian PDF of the ON duration in minutes, Gaussian PDFs are defined as $\mathcal{N}(\mu,\,\sigma^{2})$, and $\mathbf{y}$ is the power demand trace of the $\alpha_i$ from $1...T$. Note that $\mathcal{N}_{P0}$ is a Gaussian PDF of the OFF power demand as some appliances remain in standby mode and still consume power. Our current proposed UNILM solution does not use $\mathcal{N}_{P0}$; however, future algorithms will. Note that we track the minimum power ON (i.e., the power from constantly-ON appliances) as $\check{y}$.

To help choose what set of appliances are ON, we created a probabilistic version of the multiple-choice knapsack problem (MCKP)~\cite{kellerer2003knapsack}. 
MCKP is defined as:

\begin{align*}
\text{maximize } & \sum_{i=1}^m\sum_{j\in N_i} p_{ij} x_{ij} \\
\text{subject to } & \sum_{i=1}^m\sum_{j\in N_i} w_{ij} x_{ij} \leq c, \\
& \sum_{j \in N_i}x_{ij} = 1, \quad i = 1, ..., m, \\
& x_{ij} \in \{0,1\}, \quad i = 1, ..., m,  j \in N_i, 
\end{align*} 

\noindent
where 
we have $N_1,...,N_m$ items,
each item $j \in N_i$ has a profit of $p_{ij}$ and weight of $w_{ij}$,
a knapsack of capacity $c$,
and binary variable $x_{ij} = 1$ 
if and only if item $j$ is chosen in class $N_i$. Notes that MCKP notation is specific to \cite{kellerer2003knapsack} and is not used in other equations defined in our proposed UNILM algorithm.

Using the appliance models defined above we select the multiple-choice  values for each knapsack item (i.e., appliance):

\begin{equation}
    \mathbf{x}_t, p = \text{MCKP}(|\Delta y|, \mathbf{U}) , 
\end{equation}

\noindent
where $\mathbf{x}_t$ is a binary vector of length $M$ (when $\mathbf{x}_t[i]$ is $1$ then appliance $\alpha_i$ is ON, else $0$ is OFF), $p$ denotes the profit between 0--100, and $\mathbf{U}$ is a vector of length $M$ where each $\mathbf{u}_i$ is a vector of possible power values to choose from the $i$-th appliance which represent three standard deviations of integer values in $\mathcal{N}_{P1}$.

\subsection{Unsupervised Online Appliance Learning \& Tracking}

We combine the probabilistic features in Gaussian distribution $f_N()$ in the appliances models and the optimization of MCKP in Algorithm~\ref{alg:tracker} to find and track appliances online without prior knowledge using.




\begin{algorithm}
  \caption{${\mbox{\sc Track-Appliances:}}$} 
  \label{alg:tracker}
  \begin{algorithmic}[1] 
\FOR{$t \gets 2$ \TO $T$}

  \STATE $\Delta y \gets y_t - y_{t-1}$
  \STATE $\check{y} \gets y_t \text{ \textbf{when} } \check{y} > y_t$
  \STATE
  
  \IF {$\Delta y \leq -s$}
    \STATE $\mathbf{x}_t, p \gets \text{MCKP}(|\Delta y|, \mathbf{U})$
    \STATE
    
    \IF {$p > 90$}
      \STATE $\mathbf{A}[\mathbf{x}_t = 1].\text{turn\_off}()$
    \ELSE
      \STATE $\mathbf{A}[\text{argmax}_{\mathbf{A}}f_N(\Delta y,\alpha_i)].\text{turn\_off}()$
    \ENDIF

  \STATE
  \ELSIF {$\Delta y \geq +s$}
    \STATE $\mathbf{x}_t, p \gets \text{MCKP}(|\Delta y|, \mathbf{U})$
    \STATE
    
    \IF {$p > 90$}
      \STATE $\mathbf{A}[\mathbf{x}_t = 1].\text{turn\_on}()$
    \ELSIF {Mahalanobis\_distance$(\alpha_i \in \mathbf{A}) < 20$}
      \STATE $\alpha_i.\text{turn\_on}()$
    \ELSE
      \STATE $\mathbf{A}.\text{add\_new}(\Delta y)$
      \STATE $M = M + 1$
    \ENDIF
    \STATE
    
  \ENDIF
\ENDFOR
  
    \end{algorithmic}
\end{algorithm}

\subsection{Transfer Learning via Labelled Partition Maps}

Our use of partition maps can be defined as $\mathbf{P}$, a 3D matrix ($D{\times}P{\times}R$), where $D$ is the maximum duration (in minutes) an appliance can operate for, $P$ is the maximum power demand value in Watts, $R$ is the number of different regions around the world, and $\mathbf{P}[d, p, r]$ would give a specific label; e.g., ``clothes dryer''. See Figure~\ref{fig:map} for a visual representation of one region that we used in testing.

There exists a set of labels $\mathbf{L}$. Each appliance in $\mathbf{A}$ is assigned a label resulting in a binary vector $\mathbf{l_j}$ of length $M$, where $1$ indicates appliance $\alpha_i$ is assigned this label, else $0$; with the restriction that appliance $\alpha_i$ can only be assigned to one and only one label. 
We can merge appliances from our database together if they may have been assigned the same label for the partition map.

\section{Experiments}

\subsection{Experimental Setup}
We take data from House-1, Block-1, (comprised of nine day's worth of data) of the RAE dataset~\cite{makonin2018rae} which is sampled at 1Hz. 
Our disaggregator was coded in Python 3.6. We chose Python because of its array manipulation capabilities, digital signal processing libraries; moreover, its convenience when coding rapid prototypes that can get quickly translated to faster coding languages (e.g., C). All tests ran on a Mac Pro (2017 model) with a
2.3GHz Intel Core i5 processor and 8GB of memory. Furthermore, we run all our tests in one day's worth of data without any prior knowledge. 

\subsection{Experimental Results}

To preprocess/filter and disaggregate 5400 samples took a total run-time of 14.9 minutes (see Table~\ref{tbl:run}). 

\begin{table}[h]
\renewcommand{\arraystretch}{1.3}
\caption{Experimental Run-Times}
\label{tbl:run}
\centering
\begin{tabular}{|l|r|r|}
\hline
{\bf Preccess/Step} & {\bf Time (sec)}  & {\bf Time (min)} \\
\hline
1a. Median Filter           &  1.6 &  0.0 \\
1b. Bilateral Filter        &  12.7 &  0.2 \\
1c. Anisotropic Filter      &  0.1 &  0.0 \\
1d. Edge-Preserving Filter  &  875.4 &  14.6 \\
1e. Edge Sharpening         &  0.8 &  0.0 \\
\hline
1. Filter Pipeline     &  890.6 &  14.8 \\
2. Appliance Tracking  &  1.5 &  0.0 \\
3. Appliance Labelling &  0.2 &  0.0 \\
\hline
Total Run-Time      &  892.3 &  14.9 \\
\hline
\end{tabular}
\end{table}

\begin{figure}
\vspace{-0.2in}
    \centering       
        \subfigure{
           \includegraphics[width=0.98\columnwidth]{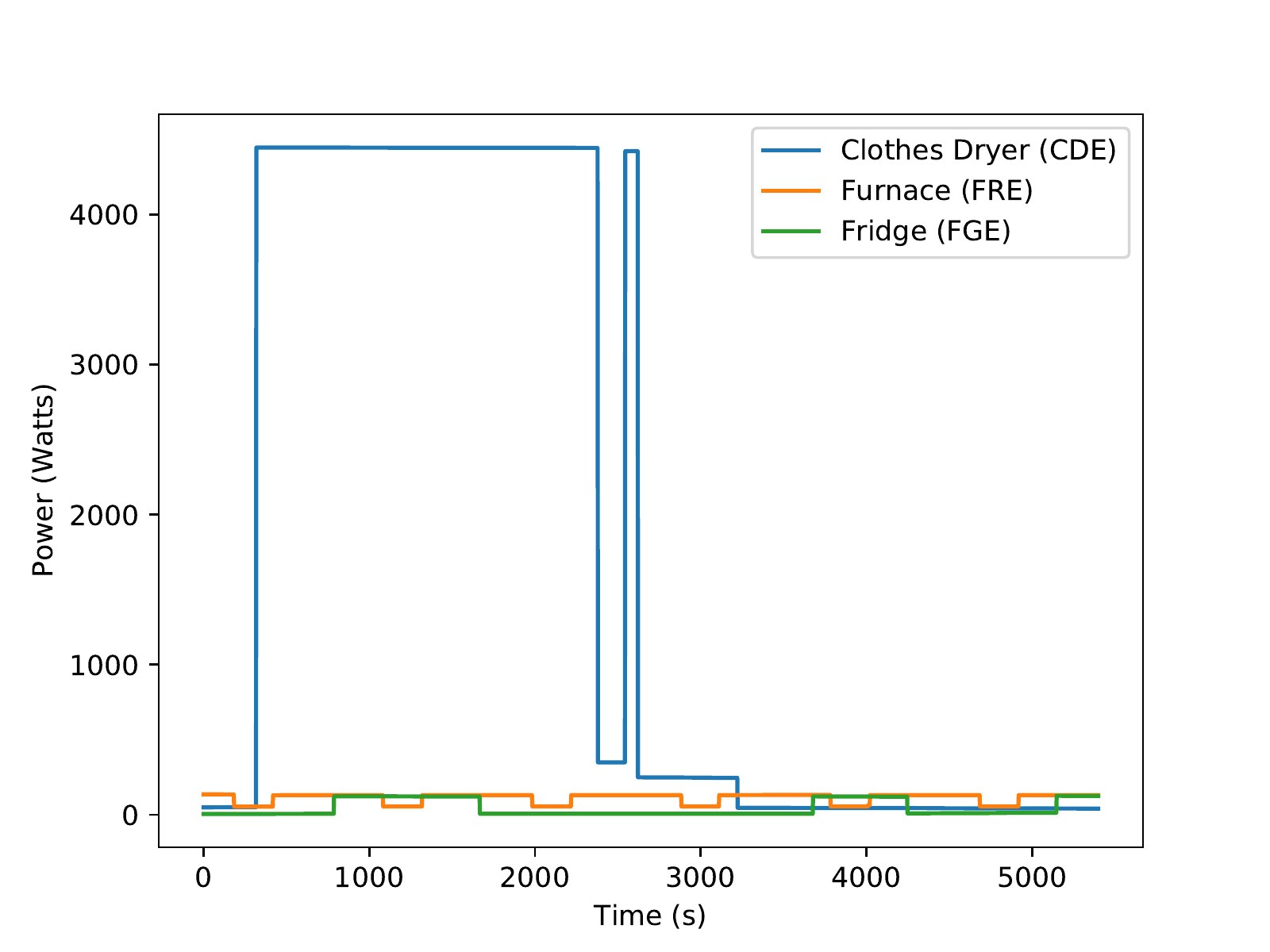}
           \label{fig:truth}
        }
        \subfigure{
            \includegraphics[width=0.98\columnwidth]{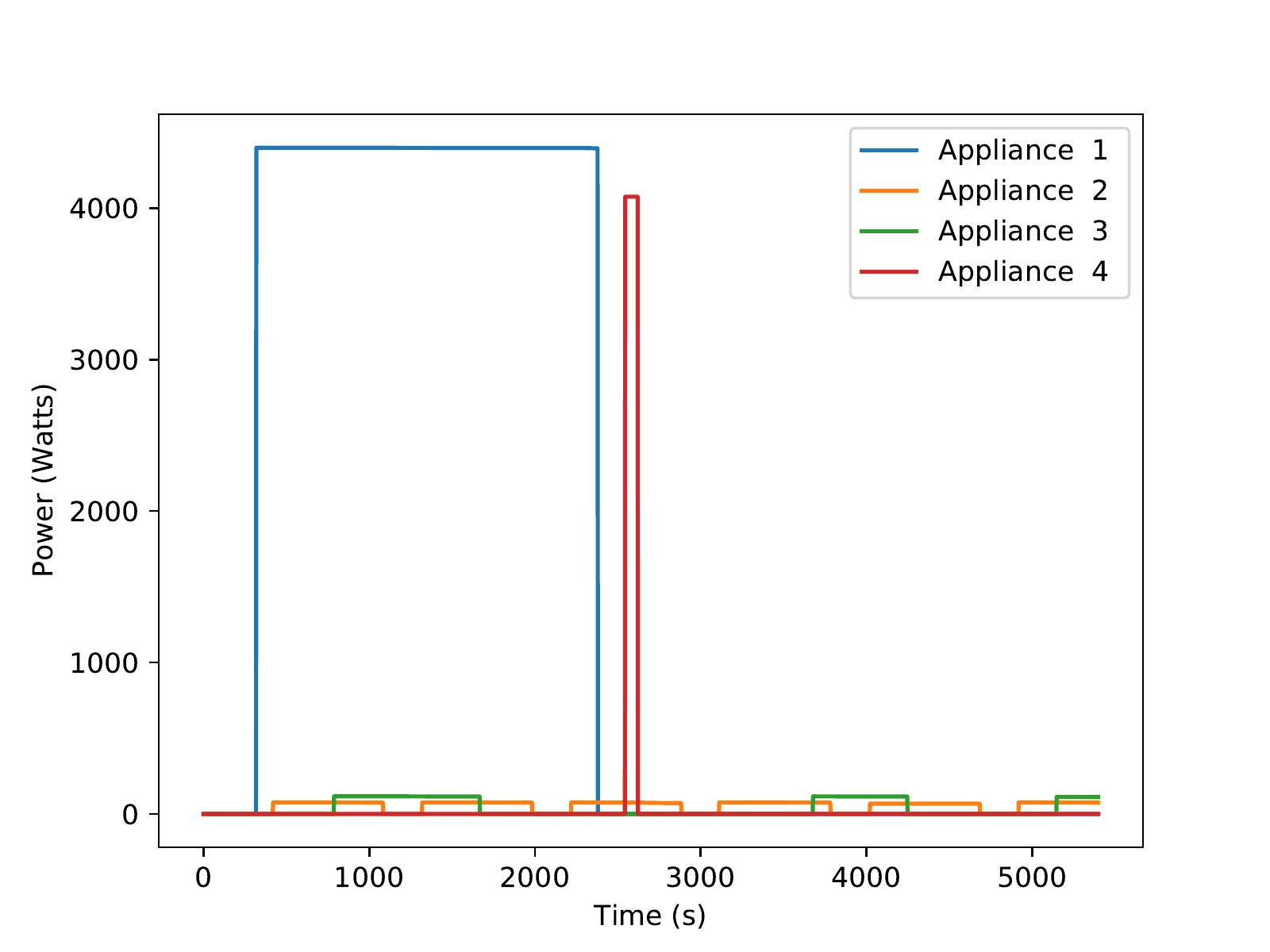}
            \label{fig:nilm}
        }
        \subfigure{
            \includegraphics[width=0.98\columnwidth]{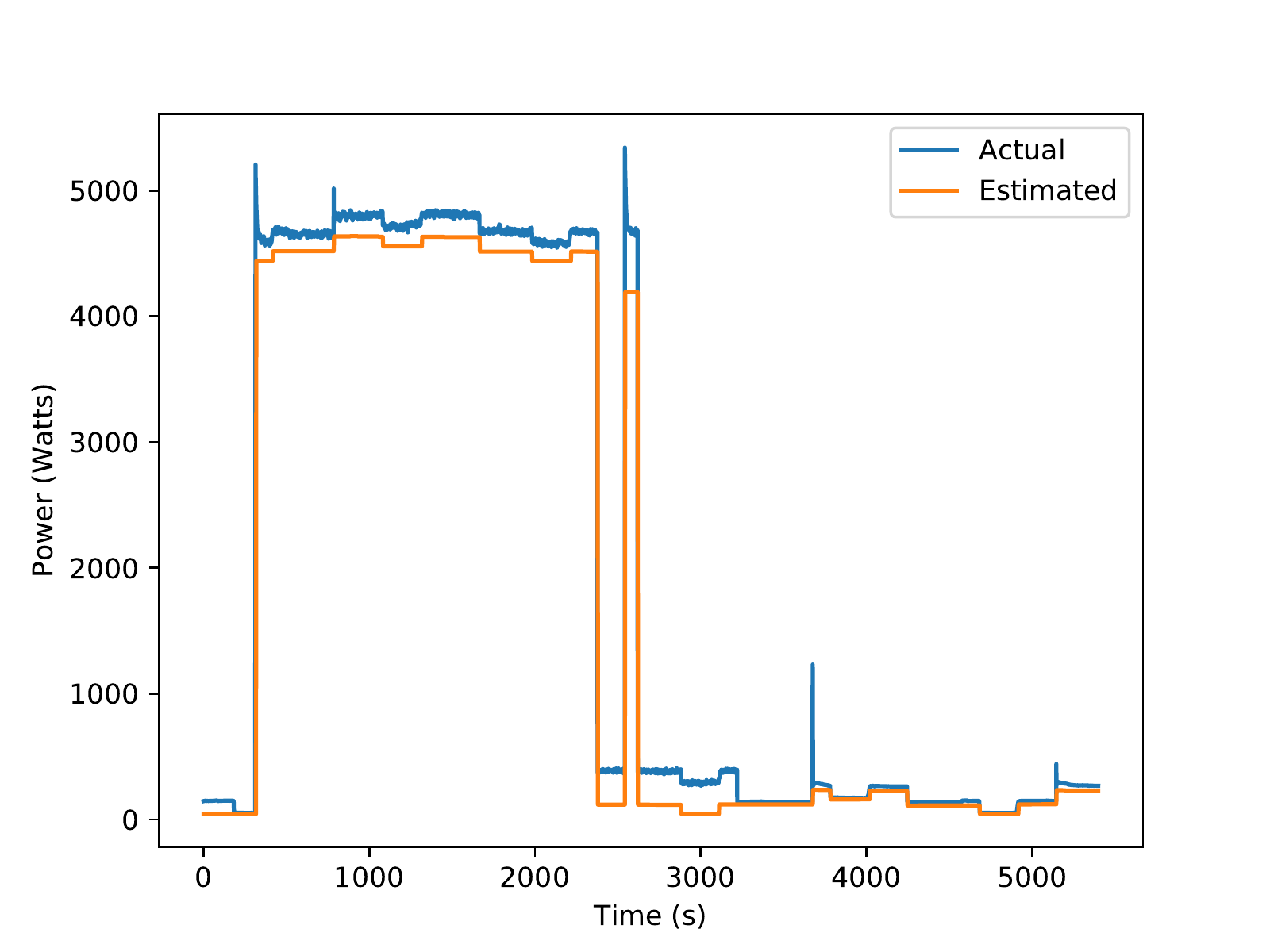}
            \label{fig:agg}
        }
    \caption{Comparing the ground truth sub-metered appliance signals (top) to that of our UNILM algorithm (middle). Ground truth aggregate to the disaggregated aggregate (bottom) showing 93.7\% accuracy in tracking.}
    \label{fig:disagg}
\end{figure}

Figure~\ref{fig:disagg} shows the individual appliances, ground truth (top) and disaggregated (middle). We can observe that the disaggregator detected four appliances (one more appliance than in the ground truth). Nevertheless, during the partition map stage, we merged these two unlabelled appliances (coloured blue and red) into one labelled appliance -- the clothes dryer -- because in the partition map they have resolve to the same label. 

Further, the output from the disaggregator (see Figure~\ref{fig:disagg}), ignored the clothes dryer from sample 2620 to sample 3220. It was not able to track the approximately 200W operational state the clothes dryer was running in. Therefore, the disaggregator ignored this OFF event. Although the system was not able to track this part of the signal, the power consumption from the clothes dryer at that interval was very small and did not considerably affect the overall accuracy score.

\begin{table}[h]
\renewcommand{\arraystretch}{1.3}
\caption{Energy Truth/Filtered/Tracked}
\label{tbl:energy}
\centering
\begin{tabular}{|l|r|r|r|r|}
\hline
{\bf Appliance} & {\bf G.Truth} & {\bf Filtered} & {\bf Est/Tracked} & {\bf Truth \textit{vs} Est} \\
\hline
Clothes Dryer & 2.753 kWh & 2.753 kWh & 2.604 kWh & 94.5\%  \\
Fridge	      & 0.063 kWh & 0.065 kWh & 0.055 kWh & 87.3\%  \\
Furnace	      & 0.174 kWh & 0.167 kWh & 0.144 kWh & 82.8\%  \\
\hline
Aggregate     & 2.990 kWh & 2.961 kWh & 2.803 kWh & 93.7\%  \\
\hline
\end{tabular}
\end{table}

To determine energy tracked, we integrate our 1Hz power (in Watts) samples to energy measured in kWh, as such:
\begin{equation}
    energy = \int_0^T \frac{y_t \times \frac{1}{3600}}{1000}\,dt .
\end{equation}

Figure~\ref{fig:disagg} (bottom) depicts both how close the total energy tracked was as compared to the raw ground truth aggregated signal. The system was able to track 93.7\% of the total aggregate energy without the use of prior information (see Table~\ref{tbl:energy}). This accuracy measure is similar \textit{normalized disaggregation error} (NDE)~\cite{kolter2012approximate,makonin2015eval}. We define the accuracy measure as:

\begin{equation}
    accuracy = \frac{energy(\mathbf{y})}{energy(\mathbf{z})} \times 100
\end{equation}

\noindent
Using the same nomenclature defined in previous sections. Some of the contributing factored to a lower accuracy for the fridge (87.3\%) was that fact that the ON-spikes from the compressor were filtered out. In reality the energy contributions from these ON-spikes is negligible, but for the small sample period in this case. Lower accuracy for the furnace (82.8\%) may have been caused by the fact that it has a constantly-ON power reading of 44W. 

        

\section{Conclusions}

We have demonstrated a prototype \textit{universal} NILM solution that can track relatively complex appliance signals without  priors. Our experiment was able to assign 93.7\% of the total aggregate energy consumed to the appliances it tracked. Although these experiments may be preliminary, they show promise for a NILM solution that works independent of country/region and without prior knowledge for modelling appliances.
Disaggregation solutions (such as UNILM) could break the economic divide allowing everyone, no matter their socioeconomic situation, to participate in energy conservation.



%

\bibliographystyle{IEEEtran}
\bibliography{refs,IEEEabrv}

\end{document}